\documentclass[12pt]{article}
\usepackage{authblk}
\usepackage[top=2cm, bottom=2cm, left=3cm, right=3cm, headheight=14.5pt]{geometry}
\usepackage{fancyhdr}
\usepackage{setspace}
\usepackage{paralist}
\usepackage{hyperref}
\usepackage{paratype}

\usepackage{graphicx} \graphicspath{ {images/} }

\usepackage{tikz}
\usetikzlibrary{positioning,fit}
\usetikzlibrary{arrows.meta}
\tikzset{tipM/.tip={Straight Barb[reversed,width=1.7ex,length=0.85ex]Butt Cap}}
\tikzset{tip1/.tip={Bar[sep=1ex,width=1.7ex]Butt Cap}}
\tikzset{box/.style={text depth=1.2ex, text height=2ex, rectangle, draw=black}}
\tikzset{tbl/.style={anchor=west, draw=black, rectangle}}
\tikzset{fld/.style={xshift=1.5ex}}
\newenvironment{fig}
	{\begin{center}\begin{tikzpicture}[node distance=0.3\textwidth]}
	{\end{tikzpicture}\end{center}}

\usepackage{changepage}

\newcommand{\bb}[1]{\textbf{#1}}

\usepackage{fancyvrb}
\DefineVerbatimEnvironment{code}{Verbatim}{fontfamily=lmtt}
\DefineVerbatimEnvironment{codeh}{Verbatim}{fontfamily=lmtt, commandchars={\\\{\}}}

\begin{document}

\title{A Next Generation Data Language Proposal}
\author[1]{Eugene Panferov}
\date{\footnotesize{Revision 2, 2021-02-20}}

\maketitle

\begin{abstract}
\noindent 
This paper analyzes the fatal drawback of the relational calculus not allowing relations to be domains of relations, and its consequences entrenched in SQL.
In order to overcome this obstacle we propose \bb{multitable index} -- an easily implementable upgrade to an existing data storage,
which enables a revolutionary change in the field of data languages -- the demotion of relational calculus and ``tables''.
We propose a new data language with \bb{pragmatic typisation} where types represent domain knowledge but not memory management.
The language handles \bb{sets of tuples} as first class data objects and supports set operation and tuple (de)composition operations as fluently as basic arith.
And it is equally suitable for building-into a general purpose language as well as querying a remote DB (thus removing the ubiquitous gap between SQL and ``application'').

\end{abstract}
\vspace{2pc}

\section{Introduction} \label{intro}

In the year 1972 E. F. Codd stated his anticipation as follows: \emph{In the near future, we can expect a great variety of languages to be proposed for interrogating and updating data bases.} It is now 2021. The only data language of any significance is SQL. Something clearly has gone wrong?

We attempt to analyze what aspects of SQL are holding back any development in the area of data languages, and propose radical solutions in both abstraction layers: data language and data storage.
Back in 1970s data languages were seen as sub-languages of general-purpose languages \cite{codd2}.
Instead something very opposite has grown -- a complete perfect isolation of general purpose languages from the database interrogation language -- splitting every program in two poorly interacting parts.
Perhaps this venue of development is ``logical'' for the language theorists (simpler formulated problems lead to better written papers of carefully limited scope that are conducive of safer grants).
But in the real life what do you need a programming language for except for DATA PROCESSING, and what do you interrogate a database for except for DATA PROCESSING.

That means the abyss between a data language (now standalone SQL) and a general purpose language has to be bridged. The procedural sub-languages were developed in the SQL environment, that could be thought of as such an attempt, that is so far failed. We approach this abyss from the opposite direction, trying to develop a brand new language that contains a native data sub-language, however the focus is on this data sub-language as this is the bloody problem.       
  
First of all, we focus on the fundamental drawback of the relational calculus: it does not allow relations to be domains of relations \cite[section 3]{codd2}. Whereas addressing this problem in the design of the language (section \ref{newpara}) provides valuable insights into possible significant improvements of underlying data storage (section \ref{mti}).

The proposed language (section \ref{lang}) attempts to overcome the following flaws of SQL:
\begin{itemize}
\item human language mimicking (which leads to inconsistent, completely inextensible, and needlessly complicated syntax);
\item mixing relational operations together and with the output formatting;
\item imperative DDL;
\item impossibility of relations between relations (which leads to non-homogeneous representation and undermines the relation abstraction itself, lowering the level of programming)
\item  separation of the data language from the application layer language (which results in the development of a complete duplicate data scheme that exists outside the relational database and suffers all the drawbacks that the relational database tried to solve in the first place, besides doubling the development labour while adding nothing)
\end{itemize}

The proposed language aims to integrate with a general purpose language used for describing an application layer, in order to keep the relational data representation throughout all processing stages as the primary and the only data-model for a given program, accessible as first-class data-objects in a general purpose language (i.e. participating as arguments in native expressions).

The proposed language abandons surrogate and foreign keys (thus making the level of abstraction higher than SQL).
The proposed data language is capable of dealing with relations between relations uniformly, by utilizing the proposed mechanism of multitable index, that in turn eliminates the need for \emph{joins}.

\section{State Of The Art}

We have no choice but to concentrate on SQL, since it is the uncontested data language nowadays.
SQL is a huge, unbelievable success. It is the most successful non-imperative algorithmic language ever existed \cite{tiobe}.
Apparently, due to the lack of competition, but the fact had to be stated.
For the same reason SQL remains basically unchanged, keeping initial imperfections and adopting only minor improvements (aka ``features'').
We witness the lack of meaningful development in the entire field of data-languages both within and outside SQL.
SQL retains all its prime imperfections, and these are not implementation imperfections nor any quantitative deficiency,
SQL contains fundamental mistakes in its very design stemming from its theoretical foundation,
which turned several pathological use-cases of the language into a necessity of life.
So much so it will be difficult to explain the severity of the pathology due to its prevalence in the population.

The apparent stagnation and growing malpractices gave rise to so called ``noSQL'' movement.
Which is a destructive cult founded on pure frustration and the lack of positive insight, of which the very name is telling.
If your data language is unsatisfactory, then by throwing it away you only solve the problems it created,
but you do not solve the problem of accessing your data, that you have BEFORE the adoption of the data language.
Nevertheless, this motion metastasized into the heart of SQL -- ISO sanctioned XML contamination \cite{iso}, effectively creating a rivaling alternative to SQL within SQL -- an apparent sign of decay.

\subsection{The Multitude Of Rivalling Representations}

A very typical approach to database design is Entity-Relation diagrams.
It is very intuitive and informative way of visualizing data structure.

\begin{fig}
\node(a)[box] {$category$};
\node(b)[box, right of=a] {$item$};
\draw[{tip1}-{tipM}] (a) -- (b);
\end{fig}

In the example above we have two entities and one relation between them.
Let us map them into a relational database.
The entities will be mapped as tables, and the relation will be mapped as a foreign key.
Note, we are doing the most typical design procedure ``by the textbook''.

\begin{fig}
\matrix(a)[draw=black, anchor=west]{\\
	\node[text width=10ex] {$category$}; \\
	\node(d)[fld] {$id$}; \\
	\node[fld] {$name$}; \\
};
\matrix(b)[draw=black, anchor=west, right=of a]{\\
	\node[text width=10ex] {$item$}; \\
	\node(c)[fld] {$cat\_id$}; \\
	\node[fld] {$id$}; \\
	\node[fld] {$name$}; \\
};
\draw[->] (c) -- (d);
\end{fig}

By the same textbook, tables are supposed to represent relations!
From the perspective of relational algebra\footnote{
or relational calculus -- at this point the distinction does not show up. Colloquially we say ``algebra'' (because the textbooks were written before SQL) but all the practical implementations of relational algebra replace it with relational calculus which is better readable in its written form}
the $item$ and $category$ are relations.
So, we have three relations: Category, Item, and the relation between them.

Why do we represent two of them properly (as it meant to be) and at the same time simulate the third one by programming a foreign key?
Is there a real need for a non-homogeneous representation?
Is the relation $item-belongs-to-category$ any worse or better than $item$ or $category$?
Can we define coherently a distinction between them?

For this particular disturbance of the representation we pay by lowering the level of programming (as in ``low-level programming'').
Foreign keys are very low-level relatively to tables.
Compare a table representing a relation versus a foreign key representing a relation:

\begin{code}
     CREATE TABLE category (
       Id    INT
     , name  TEXT );
\end{code}


Here you simply declare: these two attributes constitutes a relation.
You do not care how the association between them will be built and maintained.
You can freely manipulate pairs $(id,name)$ and apply relational operations to the relation $category$.

\begin{codeh}
     CREATE TABLE item (
       Id        INT
     , category  INT \textcolor{red}{REFERENCES} category(id)
     , name      TEXT );
\end{codeh}


Here you have created a fake attribute $item.category$.
You have defined its type which have no correspondence to the domain knowledge -- it is a surrogate type --
whereas the domain knowledge is ``item belongs to category'', it is not integer.
Then you command your RDBMS to check your input in order to keep obvious garbage out (in other words, this $REFERENCES$ directive defines a subset of the integer, making the type of this fake attribute more relevant).
Thus you have manually created a relation!
You involve yourself in the very internals of its representation.
The maintenance in principle is up to you, although the machine agrees to assist by means of $REFERENCES$ directive,
the interpretation, however, is entirely up to you -- the system does not recognize this relation as a relation.
You can not apply relational operations to it.

A \emph{join} creates a relation from existing relations.
At first glance, it looks a reasonably useful operation.
But the decades of practice reveal that a resulting relation does always pre-exist its formal ``creation'' by \emph{joining}.
Let us select something from the first example $item \longrightarrow category$:

\begin{codeh}
     SELECT category.name, item.name
       FROM category, item
       WHERE \textcolor{red}{category.id = item.category}
\end{codeh}

This ``newly created'' relation $(category.name, item.name)$ was already hidden inside our database behind the foreign key (underlined).
It is not new information created inside the database, this information had been put inside the database on purpose.
Every pair of tables you ever join, you manually MADE JOINABLE!
The \emph{join} operator merely legalizes a hidden \emph{foreign-key} relation as a first-class relation.
A join is nothing more than A CONVERSION between these two ways of representing a relation.
 
But this trivial operation comes with nontrivial cost.
First it takes a significant part of programmer's labor to make all joins precisely predictable.
A programmer must make sure all joins will result in a set of relations that are meant to be stored, and no join will ever reveal any new information.
Then a programmer is forced to codify the conversion routine itself for many relations separately.
And finally, in runtime, \emph{join} is a resource consuming operation, perhaps the heaviest operation in any typical usecase.

Multiply the runtime price for the amount of times you access those \emph{joins}, because the results are being thrown away every time.
So you need to pay the price of join every time you access such relation.
You keep converting your hidden \emph{foreign-key} relations time and again.

And to add insult to injury, those costly results are already known to you!
Your database has been designed this way that you know what comes from a join.
You pay repeatedly for something you already have and always has had!

All this time a \emph{join} solves a technicality problem that has nothing to do with the domain knowledge.
Some minor inconvenience of internal representation of relations somehow became a significant part of supposedly high-level programming and legally consumed a lion share of runtime.

Because foreign keys are not capable of representing many-to-many relations, there is another alternative representation for relations: link-tables.
It is also a low-level simulation of relations, that compromises the idea of RDBMS, and it imposes some extra cost.
Here is the simplest case of a relation many-to-many:

\begin{fig}
\node(a)[box] {$book$};
\node(b)[box, right of=a] {$genre$};
\draw[{tipM}-{tipM}] (a) -- node[above]{$book\_genre$} (b);
\end{fig}

The most recommended mainstream way to represent it in terms of tables is a link-table:

\begin{fig}
\matrix(b)[draw=black, anchor=west]{\\
	\node[text width=10ex] {$book$}; \\
	\node(bi)[fld] {$id$}; \\
	\node[fld] {$title$}; \\
	\node[fld] {$author$}; \\
	\node[fld] {$. . .$}; \\
};
\matrix(bg)[right of=b, tbl]{\\
	\node[text width=10ex] {$book\_genre$}; \\
	\node(pb)[fld] {$book\_id$}; \\
	\node(pg)[fld] {$genre\_id$}; \\
};
\matrix(g)[tbl, right=of bg]{\\
	\node[text width=10ex] {$genre$}; \\
	\node(gi)[fld] {$id$}; \\
	\node[fld] {$name$}; \\
};
\draw[->] (pb) -- (bi);
\draw[->] (pg) -- (gi);
\end{fig}

Let us select genres of a book X:

\begin{code}
     SELECT genre.name FROM book_genre, genre, book
       WHERE book.title = X
         AND book_id  = book.id
         AND genre_id = genre.id
\end{code}

Nothing unusual, all three tables properly joined according to the foreign keys provided.
But, look, $book\_genre$ is a relation on the Cartesian product of $book$ and $genre$.
We need to perform a search of the dimension $(book, genre)$.
What do we do in the select above?
We produce a Cartesian product of $book$, $genre$, and $book\_genre$ itself.
And then we perform a search of the dimension  $(book, genre)^2$ -- twice bigger search space than needed!

What have we done?! We represented a relation $book\_genre$ as THREE formally independent relations.
One proper table-relation plus two foreign-keys, that require special costly treatment to be accessed at all.
And this is formally the best practice, because it is the only way to describe such a relation in terms of relations as implemented in SQL.
There are of course other methods, say \emph{arrays}, that result in non-relational methods of querying data, that a \emph{select} does not cover.
 
Please note that in this representation the foreign keys on their own do not represent any domain knowledge relation.
It is semantically a different role of a foreign key -- not all your foreign keys correspond to a relation, and you must be aware of that --
do you have a facility in SQL to distinguish these important usecases? Ah! Sweet confusion piles up!

Some RDBMSes introduce subclasses and inheritance \cite{pgsqlinh}.
Needless to say that a subclass is a relation on a class, therefore, we have another rival representation.
How many have you counted already?
 
Some RDBMSes introduce complex types, arrays, collections\footnote{collections a.k.a. ``nested tables'' NB!} \cite{oracle} which are relation representations too.
Take a look at oracle documentation:

\begin{code}
     ---- a collection is defined as:
     CREATE OR REPLACE TYPE emp AS OBJECT (
       e_name VARCHAR2(53),
       s_sec VARCHAR2(11),
       addr VARCHAR2(113) );
     ---- You can create a table with this object
\end{code}

This is a relation, no more and no less.
So far we have SEVEN alternative representation for relations, of which only one enjoys availability of relational operations provided by so called RDBMSes.

Which representation to choose for a particular relation?
Is there any method to choose representations from this multitude?
Why so many representations after all?

The ``why'' question has an easy and obvious answer: because the notion of \emph{table} was never satisfactory for the role of a relation.
All those ``improvement'' efforts had to be strongly motivated by the need to represent relations --
should \emph{table} suffice, \emph{collections} would never be born.

\subsection{The Price Of Tables}

A typical RDBMS suggests the following mapping: $relation \rightarrow table; domain \rightarrow attribute$.
Note that a table can not adopt another table as an attribute.
Therefore, there simply is no room for relations between relations.
Taking in account that a typical domain knowledge contains a whole hierarchy of relations with the majority of them being relations between relations,
with tables we can represent only SOME of relations for any given facet of knowledge.
Once we represent a relation with a table we prevent table representations for all relations that includes the current one in their domain,
and all that are included in the current one's domain.
The representation of the former takes link-tables and foreign-keys.
The representation of the latter takes complex types and collections.

This is why we are stockpiling alternative representations.
And it is not a solution at all.
An RDBMS controls relations represented as tables, which are a mere part of a data model, which consists of many more relations having alternative representations, so that our data model resides partially (and mostly) outside a system designed for proper maintenance of the data-model.

At this point the very name ``RDBMS'' loses its meaning.

\section{Multitable Index} \label{mti}

Any facet of real life knowledge is not a plain set of relations, it is always a hierarchy of relations.
Only a minority of relations are leaves of this hierarchy.
Typically a majority of relations adopt other relations as domains.
For a meaningful system of managing a relational data model, first of all, we must make possible relations between relations.

Our language must contain the following closure (for definitions we use intuitive BNF-like notation \cite{rao}):

\begin{codeh}
     relation ::= domains graph
     domains  ::= domain
     domains  ::= domains domain
     \textcolor{red}{domain   ::= relation}
     domain   ::= scalar_type
     scalar_type // provided by an underlying system
     graph       // to be defined later
\end{codeh}

This is very important from the standpoint of the set theory that a relation (being a set) might play a role of a domain. Exactly this is missed in relational calculus and relational algebra as defined by Codd in \cite{codd2} and \cite{codd1}
The $scalar\_type$ is merely a predefined set, provided by an underlying computational system, for example $number$ or $string of characters$ were considered basic types by E. F. Codd in his original definition of the relational model of data.
Everything is plain and clear in this definition except for the mysterious $graph$\footnote{by $graph$ we literally mean the graph of a relation as in the definition of relation \cite{wikirel}} that is about to be defined.

Let's say we have a binary relation $\rho$ between sets $A$ and $B$:

\begin{fig}
\node(a)[box] {$A$};
\node(b)[box, right of=a] {$B$};
\draw[{tipM}-{tipM}] (a) -- node[above]{$\rho$} (b);
\end{fig}

\begin{fig}
\matrix(a)[anchor=west]{\\
	\node{$keys$}; \\
	\node(a1)[fld] {$1$}; \\
	\node(a4)[fld] {$4$}; \\
	\node(a7)[fld] {$7$}; \\
	\node(a9)[fld] {$9$}; \\
};
\matrix(b)[right=of a]{\\
	\node{$keys$}; \\
	\node(b2)[fld] {$2$}; \\
	\node(b3)[fld] {$3$}; \\
	\node(b8)[fld] {$8$}; \\
	\node(b9)[fld] {$9$}; \\
};
\draw[<->] (a1) -- (b2);
\draw[<->] (a1) -- (b3);
\draw[<->] (a4) -- (b8);
\draw[<->] (a4) -- (b9);
\draw[<->] (a7) -- (b3);
\draw[<->] (a9) -- node[below, color=red]{relation graph} (b9);
\end{fig}

Assuming we already have sets $A$, $B$ somehow represented, we need to represent the legs of the graph.
Apparently the relation graph's legs comprise a finite set, and any finite set can be represented as a set of integers, so we can represent our relation graph as:
$$G(\rho) = \{12, 13, 48, 49, 73, 99\}$$

Nothing is easier than creating an index for a finite set of integers:

\begin{fig}
\node(a)[box] {$\ 50\ $};
\node(b)[box, below=5ex, xshift=-10ex] {$\ 25\ $};
\node(c)[box, below=5ex, xshift=10ex] {$\ 75\ $};
\draw[-] (a) -- (b);
\draw[-] (a) -- (c);
\node(l1)[box, below=12ex, xshift=-18ex] {$\ 12; 13\ $};
\node(l2)[box, below=12ex, xshift=-4ex] {$\ 48\ $};
\draw[-] (b) -- (l1);
\draw[-] (b) -- (l2);
\node(l3)[box, below=12ex, xshift=4ex] {$\ 49; 73\ $};
\node(l4)[box, below=12ex, xshift=18ex] {$\ 99\ $};
\draw[-] (c) -- (l3);
\draw[-] (c) -- (l4);
\end{fig}

This index (by-design) represents the graph of the relation $\rho$.
At the same time it is merely a regular featureless index, like any other index in your RDBMS.
Whereas $\rho$ is a ``horrible'' many-to-many relation.
And it is being represented by a single trivial index.

From an alternative vantage point, the primary purpose of a relation graph is to answer the question: \emph{whether a tuple is a member of the relation} -- this question an index is supposed to answer too.
A graph and an index share the purpose and share representation.
For all practical intents and purposes: a relation graph IS an index.

Nothing prevents this index from holding data-storage node references.
Nothing prevents your RDBMS to maintain this index.

But we did not impose any restrictions on the domains $A$ and $B$ -- these are just sets (we merely enumerated them with integer keys) -- therefore $A$, $B$ can be relations.

Thus we have created a many-to-many relation between relations that does not require neither link-tables nor joins but a SINGLE INDEX.
An index that possesses no single feature impossible for any normal RDBMS. 

The only special feature of this index is that it has to hold two storage-node references at each leaf -- nothing more.
Generally these references point to different tables, so we called this type of index ``multitable index''.

Introduction of such index makes all the difference in the world for almost no price.
A multitable index allows to store relations between relations, treat them as first-class relations, access them without \emph{joining}
(it basically stores the information a \emph{join} re-calculates at every call).
At this point the restriction for indexes to be limited to a single table looks ridiculously artificial and unjustified.

\section{The Paradigm Shift} \label{newpara}

The language practices of SQL made \emph{relations} so strongly associated with \emph{tables} that the notion of \emph{relation} is totally replaced by \emph{table}.
You say \emph{relation} you mean \emph{table}.
The word \emph{relation} has become a geeky euphemism for \emph{tables}.
It is grossly wrong, for all the reasons exhaustively described above.

Therefore, a huge mental paradigm shift is required, in order to reflect the mathematical reality inside a properly built RDBMS:
\begin{center}\bfseries\large
Tables aren't relations! – Indexes are!
\end{center}

Indeed in PostgreSQL indexes internally ARE relations, already.
Somewhat special, but nothing should prevent you from using them as described above, since an index is capable of storing all the information you demand from a very heavy join of two tables together with a link-table.

\section{The Language} \label{lang}

We need a language to be pure, simple, and coherent.
Its notation must be clear, unambiguous, and intuitively human readable (but not mimicking a human language, in fact resemblance of a human language does only complicate understanding \cite{dijk}).
Ideally, similar objects must be described by similar sentences of the language, while dissimilar ones be described by easily distinguishable sentences.
Also, we want to keep a number of keywords and unique syntax constructs to the bare minimum.
Since everything has been already invented, we will try to stick with s-expressions \cite{sperber} and follow the functional style.

\textbf{No more human language mimicking.}
Contrary to the creators' intent, decades of human language mimicking, did not not make a single layman fluent in SQL.
This handicap should not be carried any farther.

\textbf{Relational operations to be written separately.}
We do not want to mix projection with other operation, and defer all the $limits$ and $orders$ up to the output formatting phase.
Hopefully it will make the notation of data manipulation cleaner, at the same time the output formatting richer (as being freed from the limitations imposed by the data manipulation context).  

\textbf{Declutter the notation.}
We will keep the notation free from meaningless variety of separators -- a space is enough.
For example, if we want to construct a triple, we have to provide triple members (and optionally their order), like this: $(x1, x2, x3)$.
The question is what information the comma symbol adds to this notation?
The only right answer is: void.
Because of this, we simply discard the garbage, so we got: $(x1 x2 x3)$.
This seemingly superficial change in fact is a very significant improvement to the syntax.
It effectively removes the whole idea of ``in between items'', which ordinarily causes whole series of tiny annoying problems
(particularly nasty in machine generated scripting (beginning with ``duplicate separator'' (is a duplicate separator an error? could it have a meaning?))). 

\textbf{Make basic types and relations interchangeable.}
This is the pivot point of the language.
It makes the language capable of expressing relations between relations.

\textbf{Introduce variables and assignments.}
SQL does not provide a room for assignments, they are totally alien to the SQL's structure, yet strongly demanded
(recently introduced $WITH$ clause is a counter-intuitive assignment in essence).
Our variables will be IMMUTABLE, will have a single transaction lifespan and visibility,
will be interchangeable with relations in every context except for data definition,
and will represent only sets of tuples.
Perhaps, assignments have no need to be calculated immediately.

\textbf{Make DML returning value.}
Since we have explicit output operations, and separated relational operations, and we have assignments, then we can avoid the $RETURNING$ clause altogether
by making DML return affected rows.
And since we may utilize a return value only explicitly, then we can just discard it by not utilizing it.

\textbf{Respect the fact that a relation is a function of its primary key.}
Indeed we can treat them as functions all the way long and that gives us an opportunity to create a procedural language around the data sublanguage.

\textbf{Keep types as few as possible.}
The epoch of counting bytes is over.
There is no need in keeping several different types for integers, also we do not see any high level application for bitwise operations and related stuff.
We now want computers to count their bytes (if they are concerned).
On the other hand we provide a useful tool for constructing complex types of arbitrary complexity, namely relations --
we do not need to anticipate all possible user's wishes by maintaining a library of fancy peculiar types which will be rarely known and never used
(because user's wishes always prove themselves more peculiar than our wildest anticipation).

\subsection{Type System}

The language is completely NULL-free. The only known legitimate usecase of NULLS is an optional foreign key. But our language does not have foreign keys. 

The language contains the following predefined basic types: 
$$
time,
timeinterval,
text,
int,
real,
bool
$$

$int$ is ``infinite'' as is a norm nowadays, no bytes counting, and such int effectively replaces fixed points, so we don't even bother about fixed point types.

$real$ is a floating point type, and it approximates real numbers, sometimes this distinction matters, and you need a type that is not precise. So that the minimal set of numeric types is still two very different types, which difference goes beyond hardware limitations and software implementations.

$text$ is ``infinite'' (as in PostgreSQL \cite{pgsql}) -- no more ``varchars''!!! -- limitations that are not related to a data-model itself should not be present in a data-model definition. A difference for example between $varchar(10)$ and $varchar(11)$ does not represent sufficient meaning -- it must be explained why these are different types, as opposed to different user-input-validation procedures on a client side -- such explanation does not exist. You can easily verify the insignificance of the said difference by recalling how many times a typical $varchar$ gets expanded in a typical production database per week.

$time$ is an arbitrarily rounded reference to a point in time expressed in units of currently standard calendar.
The closest comparison is the PostgreSQL's $timestamptz$ \cite{pgsqltime}, however, our $time$ is much better.
Postgres loses the timezone information and it does not respect the fact that absolutely all timestamps are always rounded.
Natural timestamp input is typically rounded coarser than an internal representation of a $timestamptz$ and the information of the user requirement for timestamp granularity is lost for Postgres.

Unlike Postgres, we preserve the timezone information, and we preserve the timestamp granularity.
Our $time$ stores the \bb{time}, the \bb{timezone} it was originally inputted in, and its \bb{granularity} --
when it is ``a year 1984'' it does not become ``1984-01-01 00:00:00.000000'', it remains ``a timestamp given to the accuracy of a whole year''.

$timeinterval$ is analogous to PostgreSQL's $interval$ \cite{pgsqltime}, but better -- it does NOT share a lexeme with text constants (see \ref{lexic}).

A calendar arithmetic should be stolen from Postgres because it is so far the only proper implementation of our calendar in the entire software industry.

In addition to these basic types one complex type could be useful enough to define him in the core language: 
$$
rational
$$
Where $rational$ is a pair $(int, int)$ treated as a number by all numeric operations.
Given our ``infinite'' integers, a rational arithmetic covers literally all the holes between $real$ and $int$, because \cite{density}.
It is not complicated computationally, roughly twice the workload of integers, plus some output simplification overhead.
And the results could potentially be presented neatly, striking balance between precision and simplicity.

Such type poses a question though: Shall we throw out $int$ completely, in favour of $rational$, since $rational$ can do everything with the same precision?

\subsubsection{Automagic: Mapping, Flattening, Tuple simplification}

The language is thoroughly relational -- all data objects in the language are\footnote{they ``are'' in the sense of their treatment, you can safely assume a scalar is a set containing a single tuple of a single element, but it does not mean we actually implement all the overhead internally} sets of tuples, all of them.
From that follow certain decisions:
\begin{itemize}
\item{The availability and the role and the appearance of Cartesian products and unions \ref{basicop}.}
\item{Automagical \bb{mapping} of functions \cite{wikimap}: a function (that is not defined as a folding function) if given a set of arguments applies to each element producing a set of results.}
\item{Automagical \bb{list flattening}, if a set is inserted into another set as an element a concatenation takes place instead, which fact is not even hidden from a user, since the only constructor of sets is the $union$ operation.
A set of sets has no meaning in a relation context -- if it does, there is a relational solution: tuples -- a set contains tuples and tuples can contain sets, no flattening is called for (see example below).}
\item{Automagical tuple \bb{depth reduction}: if a tuple contains a tuple as a single element the inner tuple is consumed the depth is reduced (see example below). Otherwise, tuples can contain tuples as elements, it is meaningful, no flattening occurs.}
\item{Operations of aggregation/grouping and their opposites should be defined. A result of aggregation/grouping is still a valid relational data object enjoying all the rights and opportunities.}
\end{itemize}

\bb{Example list flattening}

Apparently the following two sets:
\begin{code}
[1 2 3 4]
[[1 2] [3 4]]
\end{code}
contain different information, even the amount of information differs (easy check: how many permutations are allowed in each case).
But what is a semantic of this distinction?

When using these sets you have to imply some meaning into the very fact of their nestedness -- meaningless structures are not allowed.

On the other hand, a relational language gives you means to write your ideas explicitly, by extending the members of the set, for example:

\begin{code}
[
{"shots" [1 2]}
{"misses" [3 4]}
]
\end{code}
now the inner sets can not be flatten with the outer set; not technically, nor semantically; members of inner sets differ in type from the members of the outer set.

Nevertheless a meaningful analogue of flattening is still possible, we can expand this ``aggregated'' set into:

\begin{code}
[{"shots" 1} {"shots" 2} {"misses" 3} {"misses" 4}]
\end{code}

Thus both flat and compact forms are legal and all the meaning is preserved (``despite'' prohibition of sets of sets).
In this light the abstract set of sets seems a mere artefact of a language,
nothing more than a distinction between a NULL pointer and an empty array

\bb{Example depth reduction}

Apparently nested tuples can express important meaning (as structuring a datum)
\begin{code}
{"nominee" {"John Doe" 1976} 2020}
\end{code}
in this example the dates member in different tuples -- they are dates of different things, say a birth date and a nomination date.

Opposite to that, there exists meaningless configuration of nested tuples:
\begin{code}
{"John Doe"} 
{{"John Doe"}}
{{{"John Doe"}}}
\end{code}

Adding depth without adding attributes does not add any new information to the original tuple.
Therefore we perform automatic reduction of depth in this easily detectable case.
In return this simplification gives us two neat features:
\begin{itemize}
\item{We can extract (with projection) individual elements (say a scalar) from deeply nested relations, loosing the depth information (i.e. a scalar will appear as a scalar regardless of the depth it was extracted from.}
\item{We can re-use the tuple constructor for type casting, since enveloping a datum into a tuple does not increase the depth, we can re-envelop a tuple into a tuple of different type to the effect of type casting (see \ref{constructors}).}  
\end{itemize}
 
\subsection{Lexic} \label{lexic}

The language is based on the following 11 lexemes: 7 literals, 4 pairs of brackets; and 2 separators.

Spaces and linebreaks are separators. Brackets separate lexemes too.
A lexeme, unless quoted, can not contain separators or brackets.
Beginning of an operator terminates the previous lexeme.
No other separators exist.

\subsubsection{Literals}

Identifiers aka names, are used to designate types (among which are all relations (among which are all functions)).
They are classical names with lowercase initials.

\begin{code}
NAME = [a-z][a-zA-Z0-9_]*
\end{code}

Nominators (formerly known as variables) to designate named data objects.
They are classical names with uppercase initials.

\begin{code}
VAR = [A-Z][a-zA-Z0-9_]*
\end{code}

Numerals, that represent constants of numerical types.
They traditionally bear digit initials.
Underscores are allowed to group digits.

\begin{code}
NUM_DECIMAL = [0-9.][0-9._]*
NUM_BASE2TO9 = [2-9]#[0-9._]*
NUM_BASE10TO37 = [1-9][0-9]#[0-9a-zA-Z._]*
NUM_FLOAT = [0-9.][0-9.]*[Ee][+-][0-9][0-9]*

// 1984
// 2#100.000101
// 16#EF11_8766_AB00_001C
// .5e-189
// it is debatable if we should allow the leading "."
// mandating a leading 0 would FREE the "." for some other purpose. 
\end{code}

Text literals in double quotes with traditional backslash escapes.

\begin{code}
TEXT = "[^"]*" // with backslash escapes (not shown here)
\end{code}

Date, time, and interval literals, in ``graves'' without escaping.
Easily distinguishable from both numerals and texts by the leading character.
The use of a quotation technique allows for spaces to be used in date-time formats. 

\begin{code}
TIME = `[^`]*`

// `1984` // it is now a year not an integer
// `456BC`
// `+ 2days 7hours 11minutes`
// `2021-02-20 18:41`
\end{code}

The temptation was strong to avoid keywords completely, still there are two words that are absolutely reserved forever and ever:

\begin{code}
BOOL = true | false
\end{code}

Operators are brief cryptic names for frequently used functions.
They must have a non-alphanumeric initial that is not a quote-mark nor a bracket, they might contain letters, but not digits.
This distinctive shape allows them to appear on a non-leading position in a s-expression, without being confused with arguments.

\begin{code}
OPERATOR = [!@#\$\%\^\&\?~*+=<>:;,|\\/-][a-zA-Z_!@#\$\%\^\&\?~*+=<>:;,|\\/-]*
\end{code}

\subsection{Syntax}

In order to focus on the data sublanguage we omit for now the definition of a program in the proposed language.
Defining that, including the command grouping and structuring into transactions and routines would be premature at this point, since the big picture of the workflow is not complete yet.

On the other hand we attempt to define as clear as possible how does the language specify data structures and data access.
Keeping in mind that all these ``constructs'' should be equally applicable to remote and local data, and equally palatable for a language with shared state data objects, as well as a language without state sharing.
 
Typographic Convention:
\\*We will type literal value of terminals in "doublequotes".
\\*We will type names of terminals UPPERCASE, if their value is variable.
\\*We will give lowercase names for non-terminals.
\\*For brevity of lists definitions we extend our BNF-like notation with symbols: \verb|::=...| and \verb|::=0...| meaning arbitrary repetition of right-side, at least 1 or at least 0 times respectively.

\subsubsection{Data Definition}

\begin{code}
definition ::= "relation" { NAME domains }
definition ::= "domain"   { NAME domains }
domains    ::=... domain
domain     ::= NAME:type   // you can give names to domains
domain     ::= type        // domains may bear type-names
type       ::= BASIC_TYPE_NAME
type       ::= RELATION_NAME
\end{code}

There are three classes of relations: ordinary $relation$, $domain$, $function$ -- all of them are first-class relations in every respect.
A $domain$ always contains all its possible tuples, similarly to a user-defined type in procedural languages,
although from our vantage point a $relation$ is also a user-defined data type of equal rights and opportunities,
the only difference is that $relation$ represents a user-defined subset of its type (i.e. requires a set of tuples to be $added$)
and contains only those tuples that were $added$ by a user.

A $function$ is a relation that contains only tuples defined by the given expression (i.e. calculated).
The additional difference is that you can explicitly specify its co-domain to be some arbitrary projection of the complete domain --
i.e. a function represents a ``shorter'' type\footnote{The explicit split of the formal ``relation domain'' into ``argument type'' and ``result type'' also reflects the fact that in general a function specified as a computation procedure can not be reversed, thus making regular ``selects'' (potentially retrieving values of ``arguments'') impossible, whereas the ``result'' alone retains full ``selectability''.}, similarly to ``return type'' of a function in a procedural language.
For the sake of completion, here is how we define functions:

\begin{code}
definition ::= "function" { NAME domains -> domains } heads
definition ::= "fold"     { NAME domains -> domains / init} heads
definition ::= "operator" { OPERATOR FUNC_NAME }
heads     // to be defined later
init      // to be defined later
\end{code}

A $function$ that receives a set of its arguments automatically ``maps'' itself, producing a set of results.
To cover the opposite  behaviour a $fold$ is introduced.
A ``folding'' function (although enjoying equal rights and opportunities) if given a set of arguments always folds this set and produces a single result.  
$Folds$ are to be used in aggregates.
The $init$ condition contains a definition and an initial value of the accumulator that passes the fold state between iterations.
It is not supposed to be accessible from a call to a $fold$, a $fold$ participates in a program as a $function$ with the specified domain and co-domain.

An open question is: (a) do we make a function's domain type accessible by a name derived from the function name? (b) do we require explicit names for the domain and co-domain types to be defined prior to the function definition, so that a function always refers to named types\footnote{For an example usecase of type names see \ref{constructors}}?

\textbf{Example}

Let there be a tiny library:

\begin{fig}
\node(g)[box] {$genre$};
\node(b)[box, right=of g] {$book$};
\node(a)[box, right=of b] {$writer$};
\draw[{tipM}-{tipM}] (g) -- node[above]{$book\_genre$} (b);
\draw[{tipM}-{tipM}] (a) -- node[above]{$author$} (b);
\node(d)[box, yshift=-8ex, xshift=6ex] {$department$};
\draw[{tipM}-{tip1}] (b) |- node[above, pos=0.75]{$available$} (d);
\end{fig}

The definition of this library will look like:

\begin{code}
domain {name first:text middle:text last:text}
relation {book title:text ISBN:text}
relation {writer name:text birthdate:time}
relation {genre text}
relation {department text}
relation {author book writer}
relation {book_genre book genre}
relation {available book department}
\end{code}

Now let us illustrate another usecase for $domain$ definition:

\begin{code}
domain {point2d x:real y:real}
domain {circle radius:real center:point2d}
relation {my_circle circle}
\end{code}

In the example above circle is a set of all possible circles.
Domains play a role of complex types.
While my\_circle is a set of user defined circles –- a user himself controls which tuples belong to the set.

\subsubsection{Constructors} \label{constructors}

We construct a tuple simply by listing all its components in special brackets designating this is a tuple. 

\begin{code}
constructor ::= { tuple }
tuple ::=... value
value ::= expression
value ::= label:expression
label ::= TYPE_NAME
label ::= NAME
expression // to be defined later
\end{code}

Tuples are allowed to have explicitly or implicitly named elements -- exactly the same as in a definition of a relation.
Elements will be referred to by their labels, either arbitrary names or their respective type names, for the purpose of unification\footnote{in principle $unification$ as in Prolog and Erlang, but greatly enhanced}.
This allows us to abandon all traces of positional referencing in all contexts.
Tuples are still ordered but this order only manifests in its useful aspect: establishing correspondence between elements of two tuples, and this is achieved with the aid of $labels$ -- the rest of the information that an order of elements contains is not important for a user, and removed out of sight.

In addition to that, you can ensure ``compatibility'' of a newly constructed tuple with a predefined type:

\begin{code} 
constructor ::= { TYPE_NAME tuple }
\end{code}

If the given tuple fails to unify with the given type the constructor returns an error.
Apart from type assurance this form adds human readability in certain cases.

On top of it (due to automatic depth reduction) you can use the same type assuring tuple constructor for type casts:

\begin{code}
A := {int 1984}
B := {text A}      // result "1984"
C := {time B} // result `1984`
\end{code}
 
The second necessary constructor is a set.
We construct a set by simply listing its elements in brackets designating a declaration of a set.

\begin{code}
constructor ::= [ set ]
set ::=0... expression
\end{code}

Our sets are relations, so that they imply that all their elements are of the same type.
In order to make this requirement more clearly visible we allow type assurance for sets as well.
This constructor will apply the type assurance to each element of the given set:

\begin{code} 
constructor ::= [ TYPE_NAME set ]
\end{code}

\bb{Counter-example}

Let us highlight the importance of explicit tuple constructors by looking at SQL, which is built around tuples but lacks syntax for constructing tuples.
This situation leads to queries like that:

\begin{codeh}
SELECT sum(salary) \textcolor{red}{OVER w}, avg(salary) \textcolor{red}{OVER w}
   FROM empsalary
   WINDOW w AS (PARTITION BY depname ORDER BY salary);
\end{codeh}

it says:
\begin{code}
x over w, y over w, z over w
\end{code}

because there is no way to say:
\begin{code}
{x y z} over w
\end{code}

\subsubsection{Basic Set Operations} \label{basicop}

A relational language can not survive without the basic set operations: Cartesian product and union -- being as easily available as arithmetic operations.

\begin{code}
product ::= { expressions }
union ::= [ expressions ]
expressions ::=... expression
\end{code}

We use the same brackets here, respectively. A \verb|[]| designates a set. A \verb|{}| designates a tuple.
It must be obvious from the definition that the similarity goes beyond the syntax:
a tuple constructor performs Cartesian product of its elements,
and a set constructor performs a union of its elements -- there is no need so far to invent any distinctions.
Of course you can specify type assurance too.

\subsubsection{Unification}

The type assurance, and function calls, and the pattern matching in selections (see \ref{selections}), are based on $unification$.
Our unification procedure utilizes both labels and types of the tuple, ignoring the order, and in some cases ignoring arity.

\begin{code}
domain {name first:text middle:text last:text}
// unifies with:
{first:"John" middle:"Enoch" last:"Doe"}
{first:"John" last:"Enoch" middle:"E."}
{first:"John" middle:"Enoch" "Doe"} // the last is implied
{"Doe" first:"John" middle:"Enoch"} // the last is still implied

domain {point x:real y:real name:text}
// unifies with
{x:10 {real "19"} "point 10;19"} // type matters
{x:"10" y:"19" "point 10;19"} // but labels take precedence

// abbreviations are allowed
{name f:"John" l:"Doe"} // is valid
// as long as abbreviations are unambiguous
\end{code}
 
Label abbreviations are possible because of the very narrow scope of the labels, ambiguity is easily detectable.
Since labels are used profusely, this feature becomes extremely time-saving.
At the same time it allows to avoid abbreviating ``field names'' at the definition phase.
You can afford names as verbose as necessary without cluttering the text of a program referring to verbose names.

\bb{The unification algorithm}

Elements are consumed in the order from the highest to lowest priority:
\\* Label and Type match exactly
\\* Label match exactly and type casts
\\* Label abbreviates and Type matches
\\* Label abbreviates and Type casts
\\* Label empty and Type matches to a single element
\\* Label empty and last element and Type casts to the last remaining element.

Error is thrown:
\\* more elements in the given pattern than in the target
\\* a label in the pattern does not match nor abbreviate any target
\\* a label matches and the type fails to cast

\subsubsection{Nominators}

Giving names to subselects is a massive convenience, so much so SQL is developing this aspect for its entire history up to the ``WITH'' control structure.
We decide from the start to structure the language is such a manner that a sequence of ``assignments'' is legal without any special tricks excusing such ``assignments'' from a context that prohibits them.

\begin{code}
assignment ::= NOMINATOR := expression
\end{code}

\bb{Example}

\begin{code}
MYlist := [1 9 8 17]
MYtuple := {"John" "Doe"}
B := (book title:"1984") // B now represents this selection
\end{code}

After these assignments (within their visibility scope) you can treat the nominators $MYlist$, $MYtuple$, $B$
as the objects they represent and put them in any appropriate context.

An assignment sentence DOES \bb{NOT} necessitate execution of the right-side expression!
The compiler knows if and when the assignment is used,
keeping in mind the transactional context of the runtime,
the necessity to execute anything occurs at well defined moments.

For purely aesthetic purposes, in case programs in our language eventually grow a habit for large multiline expressions,
we must keep in mind the possibility to add postfix assignments:

\begin{code}
postfix_assignment ::= expression =: NOMINATOR
\end{code}

In a lengthy statement such assignment allows to put the $NOMINATOR$ on the very last line where it is most relevant for reading the text right below.

\begin{code}
case of
  a very lengthy
	expression
	that itself
	could be a control structure
	that returns a result
	and you want this result to be named
	and by the time
	you finished reading
	this expression
	the variable name it is assigned to
	goes far above...
	INSTEAD
	you can provide this name 
	right AFTER the present expression
	exactly at the moment 
	when you finished learning
	the result of this expression
	and so timely you learn the name too
end-case =: MYresult 

B := (foo MYresult)
\end{code}

\subsubsection{Selection} \label{selections}

When you write:

\begin{code}
SELECT * FROM book;
\end{code}

What information exactly does this sentence contain?
The right answer: $book$ -- nothing else.
In this sentence you point to a certain relation by name
(the result could be directed to the output or another select, but this information is not contained within the sentence itself)

Since we know the information content of this sentence, we know how to say it in our language:

\begin{code}
(book)
\end{code}

This is how you access a relation -- you simply call it by its name.
Parenthesis here makes it an expression that is a data object as opposed to a type object.

\begin{code}
selection ::= ( NAME )
selection ::= ( NAME pattern )
pattern ::=... filter
filter ::= conjunction
filter ::= disjunction
filter ::= value
filter ::= func
value ::= LABEL:expression
value ::= expression
func ::= LABEL:curry
curry ::= (FUNCTION args)
args ::=... value
conjunction ::= { pattern }
disjunction ::= [ pattern ]
\end{code}

A pattern attempts to match against the elements of the relation.
If a value is provided it matches with the corresponding field iff equal.
If a function is provided it matches if computes to true, but this function must be curried down to the arity 1, and return boolean result.
A set of filters matches as ``OR''. Whereas a tuple of filters matches as ``AND''.

\bb{example}

\begin{code}
Dawkins := (writer {first:"Richard" last:"Dawkins"})
John_OR_Mary := (person first:["John" "Mary"])
Elder_J_OR_M := (John_OR_Mary birthdate:(earlier than:`1990`))
// same as
Elder_J_OR_M := (person first:["John" "Mary"] birthdate:(earlier than:`1990`))

X := (person [{first:"John" birthdate:(earlier than:`1990`)} {first:"Mary"}])
//same as
X := [(person first:"John" birthdate:(earlier than:`1990`)) (person first:"Mary")]

// provided a predefined BIF
function {earlier x:time than:time -> bool} ( ( x - than ) < 0 )
\end{code} 

A $pattern$ must represent a volume in the domain space of the given relation.
A selection expression returns all the elements that happen inside this volume.
The simplest form of a $pattern$ (apart from the empty pattern) is a tuple.
A tuple-pattern unifies with some elements of relation, these elements will be returned.
A complete tuple of the relation-type represent a point in the domain space and unifies with 1 or 0 elements.
A tuple of a single element represents a hyperplane.
A tuple of multiple elements represents an intersection of hyperplanes.
Whereas a set within a pattern represents a disjunctive match (analog of SQL's: ``OR``, ``IN'').

An important class of shapes can be expressed by means of adding up and intersecting such volumes -- those boundaries can utilize indexes.
It is huge class covering almost all usecases of a typical bureaucratic database,
However, it is far from being complete, for example this selection can not express a half plane on a real plane that is bounded by $x+y=const$.
This problem is still solvable, because there is no compulsion to put everything inside a ``where'' clause.
You can define a function that computes $x+y=const$ (or any other hyperplane for that matter) taking the entire relation as an argument.

\begin{code}
relation {plot x:real y:real}
function {foo p.plot -> plot c:real} {p (+ p.x p.y)}

F := (foo (plot))
HALFPLOT := (F c:(greater than:0))
\end{code}

\subsubsection{Data Management}

\begin{code} 
command ::= "add" content
command ::= "remove" content
command ::= "abolish" content
command ::= "update" content replacement
content ::= constructor
content ::= selection
replacement // to be defined later
\end{code}
 
This weird triviality comes from the fact that a constructor specifies type -- so that the relation to which the command is issued is known from the type.
In case a given tuple (say, passed from some previous routine) is not labeled with a type,
it is always possible to do inline by wrapping it into a type-assuring constructor.
Therefore the syntax of the command itself does not need to mention the type, it is either known or can be made known by treating the argument of the command.

And of course, a set is acceptable argument.

And of course, a command returns a result: the affected subset of the affected relation.
 
\begin{code}
B := add {book title:"1984" isbn:"094885858"}
G := add [genre {"fiction"} {"sci-fi"} {"memoir"} {"documentary"} {"bore"}]
PEOPLE := [{first:"George" last:"Orwell"} {first:"Richard" last:"Dawkins"}]
W := add {writer PEOPLE} // adds all people as writers
A := add {author B (W last:"Orwell")}
S := add {book title:"The Selfish Gene" isbn:"098765"}
\end{code}

Similarly, a $remove$ command needs to know ``which records'', and since there is no preference as to how exactly they must be specified,
as long as the provided datum unifies with the relation...
The most straightforward way is to specify a selection.

\begin{code}
TRACE := remove (author (name [first "Mary" "Jane"]))
// removes all authors with first names "Mary" OR "Jane"
\end{code}

$abolish$ is a recursive version of $remove$.
It removes members of relations referring to removed members, instead of throwing an error.

The $update$ is exceptionally cluttered and burdensome with all its references to $new$ and $old$.
By no means the proposed syntax is final. Please consider it as a suggestion.

\begin{code}
command ::= "update" content replacement
replacement ::= constructor
replacement ::= expression_with_special_namespace
\end{code}

\bb{Example}

\begin{code}
update (writer {name last:"Dawwkins"}) {name last:"Dawkins"} 
// only the unified elements get replaced

// referring "old"
update (writer birth:(earlier than:1500)) {writer birth:(+ `13 days` <birth old>)} 
\end{code}

\subsubsection{Projections}

Projections are also brackets. In the simplest case:

\begin{code}
projection ::= < fields expression >
fields ::=... field
field ::= NAME
// where expression is typically a selection 
\end{code} 

Unlike other brackets, unlike selections themselves, the subject expression enters a projection at the last position.
Is it inconsistency? No!
In a selection the first word within its bracket designates the type of the result.
Same with constructors, the type enters in the first position.
A projection effectively alters the type of the subject expression, and the formulation of this new type is written in the $fields$ clause.
So that in the proposed syntax of projections the type again is the leading member within the bracket.
That adds some regularity for the reader, a human reader.

\begin{code}
B := (book title:"1984")
W := <writer (author book:B)>
N := <name W> // the type is now {name first middle last}   

T := {x:10 y:19 name:"point" ref:`1546-12-12AD`}
P := <x y name T> // dropped the "ref" field

// nesting
F := <first <name <writer (author book:B)>>>

// more complex
K := {name f:"J" m:"J" l:"Doe"}
Q := {x:10 y:19 who:K} // type is {real real {text text text}}
// let's extract x and m (from name)
Mpart := <m <who Q>>
Xpart := <x Q>
R := {x:Xpart m:Mpart}
// pretty complicated
\end{code}

In order to fight this complication it is enough to allow the $field$ specifications within projections to refer to subtype labels:

\begin{code}  
projection ::= < fields expression >
fields ::=... field
field ::= NAME
field ::= NAME.field
// where expression is typically a selection 
\end{code} 

So the last example (that explicitly flattens a nested-tuple type) will look like:

\begin{code}
K := {name f:"J" m:"J" l:"Doe"}
Q := {x:10 y:19 who:K} 
R := <x name.m Q>
\end{code}

There is nothing wrong with the dot-notation for nested namespaces.
Also we allow abbreviations in this context too.

\subsubsection{Grouping And Aggregation}

If you aggregate something in SQL and forget to ``GROUP'' unaggregated fields,
it will very precisely report an error: ``the fields x y z are not aggregated and should be grouped''.
It knows exactly what fields must be grouped!!!
Therefore, we must pursue some aggregation and grouping notation that does not demand this redundancy from a user.
 
It turns out, grouping snugs comfortably into the projection context.

\begin{code}
grouping ::= < fields \ fields expression >
\end{code}

We simply specify two subsets of fields: first those being grouped, then those that group them.
From the information content perspective, we specify a subset of fields (same as in projection) plus a single \emph{split} among them.
The question which part of the slit should come first in the notation is open. 

\begin{code}
domain {my_book title:text year:int author:writer}
[my_book
{t:"a book 1" 2021 (writer "Dawkins")}
{t:"a book 4" 2024 (writer "Dawkins")}
{t:"a book 5" 2025 (writer "Orwell")}
] =: B
G := <title year \ author.name.last B>
// [
//  { [{t:"a book 1" 2021}
//     {t:"a book 1" 2024}
//    ] "Dawkins"
//  }
//  { [{t:"a book 5" 2025}
//    ] "Orwell"
//  }
// ]
\end{code}

Since we successfully grouped certain projections into sets, and we already have folding functions,
then an aggregation is merely an application of a folding function to certain elements of a grouping result.
It can be done by means of regular function calls and does not require any special syntax.

\subsubsection{Connections} \label{connections}

Instead of useless joins, we offer so called $connections$ --
a way of writing a selection from a whole hierarchy of relations, without explicitly mentioning each of them.
The opportunity stems from the fact that the data scheme already contains all the information
how relations relate to each other, how they are connected.

\begin{code}
connection ::= ( RELATION -><- selection )
\end{code}

It results in a subset of the given relation, but the selection criteria are formulated for another relation (implied in the given selection). 
The system scans the data scheme to find a path between two relations involved.
And then connects the elements of the given selection with the elements of the queried selection along the found path.

\begin{code}
W := (writer -><- (genre "bore"))
// all writers connected to the given genre
// same as
G := (genre "bore")
B := <book (book_genre G)> // projecting book_genre on book
A := (author B) // author is a relation between book and writer
W := <writer A>
\end{code}

\subsubsection{Expressions}

Finally. For a good closure.

\begin{code}
expression ::= CONST
expression ::= NOMINATOR
expression ::= selection
expression ::= projection
expression ::= connection
expression ::= constructor
expression ::= arith
expression ::= funcall
\end{code}

\subsubsection{Composite Types}

On one hand, having strictly typed functions makes perfect sense in a language where all data objects are strictly typed.
When we query a book:
\begin{code}
(book title:"1984")
\end{code}
we know the record type we got --
the same knowledge of the type when querying a function is convenient.
On the other hand a function does not always return a result, could be an error.
Setting aside the exceptions (which remain an open question), we still need some means of handling non-exceptional results...

Since our functions return tuples, it is always sufficient to return a type such as
\begin{code}
domain {foo_result result:int error:text}
\end{code}
Although being sufficient and very convenient in certain cases of error-prone function calls,
such as calls to a filesystem that potentially return plentiful non-exceptional errors,
this solution entails the burden of repetitive projections,
e.g. always extracting the actual result from a return-tuple
\begin{code}
<result (foo arg)>
\end{code}
a return-tuple can not be used in its raw form.

For example Erlang (despite the capability of returning tuples!) utilizes untyped functions for returning error values:
a function can return a result of some known type or an error of another distinct type, the type of the result tells the fact of error.

Without forcing our users to adopt either strategy we may make both of them implementable in our language,
by introducing \emph{composite types}, a type that represents a union of several other types.
Such a type represents a compromise between strict typisation and the Erlangish approach.
Such a disjunctive type remains strict, it predicts exactly the variety of expected results, still it allows to transcend pre-existing types.

\begin{code}
type ::= "domain" {NAME domains}     // a tuple of its domains
type ::= "domain" {NAME [ domains ]} // either one of the domains
domains ::=... domain
domain ::= TYPE
domain ::= LABEL:TYPE
\end{code}

For example a value of the type $point$ can be either a 2D or 3D point:

\begin{code}
domain {point2d x:real y:real}
domain {point3d x:real y:real z:real}
domain {point [point2d point3d]}
\end{code}

In a less abstract venue, it allows us cater to some very practical usecases:

\begin{code}
domain {birthdate [date text]} // date OR text
relation {person name:text birthdate}
add {person "John Doe" `1976-01-01`}
add {person "Ali ibn Hattab" "the user refused to reveal his birthdate"}
\end{code}

Although this example puts another important nail in the NULL's coffin,
still more practical usecases are possible: 

\begin{code}
relation {prose_genre text}
relation {poetry_genre text}
domain {literature_genre [prose_genre poetry genre]}
\end{code}

This example strikingly resembles PostgreSQL's inheritance \cite{pgsqlinh}.
Although in Postgres it is defined in reverse order, a supertype first.
You can clearly see that a composite type can be a supertype in the exact same sense as in Postgres.
And thus it covers all the legitimate usecases for Postgres's inheritance and subtypes.
And we only assumed here that Postgres along with Oracle share some strong meaningful real-life motivation to invent supertypes and squeeze them into their existing RDBMs.

\section{In The Conclusion's Stead}

We have got a world's first language with \bb{pragmatic typisation} -- the types represent your domain knowledge, and refuse to represent memory management --
a choice of a type is solely dictated by a given problem but not the implementation of the data storage.

The language handles \bb{sets of tuples} as first class data objects (as ``variables'' in archaic terms) and supports set operation as fluently as arithmetic operations.
The operations of tuple composition and decomposition can be expressed in the same manner (basic arith). And to achieve that we cleansed the language from any traces of positional referencing -- all tuple entries are named.

Let's summarize the secondary features of the language so far described.
\begin{itemize}
\item{Full utilization of the information provided by the data scheme (see $connections$ \ref{connections})}
\item{Concise formulations of queries}
\item{Textual compatibility of relational data manipulations with general purpose functional languages}
\item{Higher-order relations not requiring any special treatment, nor causing any confusion}
\item{Comfortable and straightforward laziness}
\item{Bare minimum of syntax constructs: re-using and re-using ``homomorphic'' structures for ``homomorphic'' purposes}
\item{Bare minimum of basic types}
\item{Comfortable handling of types of arbitrary complexity}
\item{Named parameters, free from positional references, name abbreviations}
\item{Same full toolbox for querying and constructing ``Local'' and ``Temp'' data, using the predefined data scheme}
\item{The SQL's basic operations decomposed into more basic parts and reorganized more logically}
\item{Several improvements were made in general readability (see postfix assignments)}
\end{itemize}

Many questions remain open. Is there a way to economize bracket types and reduce the zoo by one animal?
Perhaps a bracket role can be unambiguously told by the first member class (e.g. a type as opposed to a nominator, etc)?
There is room for ambiguity in unification of patterns containing curried functions.
We have not defined any control structures, not even decided about grouping of commands.
There must be subprograms, call them procedures or routines, or ``saved transactions'' if you will, as opposed to already defined functions that are pure.
The type of the connection operation is not yet settled, it could be anything from the type of the queried relation to the Cartesian product of all types along the connecting path.

\end{document}